\documentclass[epj]{svjour}
\usepackage{graphics}
\usepackage{amssymb,amsmath}
\begin{document}
\title{RVB description of the low-energy singlets of the spin 1/2 {\it
kagom{\'e}} antiferromagnet}
\author{M.~Mambrini\inst{1} \and F.~Mila\inst{1}
}                     
\institute{Laboratoire de Physique 
Quantique, Universit\'e Paul Sabatier, 118 Route de Narbonne, 31062 
Toulouse Cedex, France. }
\date{Received: 06/03/2000 / Revised version: 06/03/2000}
%
\abstract{Extensive calculations in the short-range RVB (Resonating
valence bond)
subspace on both the trimerized and the regular (non-trimerized) Heisenberg model on the
{\it kagom\'e} lattice show that short-range dimer singlets capture
the specific low-energy features of both models. In the trimerized case the
singlet spectrum splits into bands in which the average number of
dimers lying on one type of bonds is fixed. These results are in good agreement with  
the mean field solution of an effective
model recently introduced. For the regular model one gets
a continuous, gapless spectrum, 
in qualitative agreement with exact diagonalization results.
} 
\maketitle
\section{Introduction}
\label{intro}
It is well known that the conventional picture of a long-range, ordered,
dressed N\'eel ground state (GS) can collapse for low dimensional frustrated
antiferromagnets. The GS of several spin 1/2 strongly frustrated
systems has no long range antiferromagnetic order and is separated from the
first magnetic ($S=1$) excitations by a gap. The first example
of such a behavior was given by the zigzag chain at the Majumdar-Ghosh point \cite{Majumdar}
($J_{2}/J_{1}=1/2$) in which case the two-fold degenerate GS is a product of
singlets built on the strong bonds.

In some cases the consequences of frustration on the
structure of the spectrum can be even more dramatic. It is now firmly
established by many numerical studies that the singlet-triplet gap of the
Heisenberg model on the {\it kagom\'e} lattice is filled with an exponential
number of singlet states \cite{Waldtmann}. This property is actually not specific to the
{\it kagom\'e} antiferromagnet (KAF) and could be a generic feature of strongly frustrated magnets: It is
suspected to occur also for the Heisenberg model on the pyrochlore 
lattice \cite{Canalspriv}, and it has been explicitly proved for a 
one-dimensional system of
coupled tetrahedra which can be seen as a 1D analog of pyrochlore
\cite{Mambrini}. 

Since the particular low-temperature dependence of many physical quantities
is directly connected to the structure of this non-magnetic part of the
spectrum, many recent works \cite{Waldtmann,Zengelser1,Singhhuse,Leungelser,Zengelser2,Sindzingre,Nakamura,Lecheminant} were devoted to understand the nature of the
disordered GS and low-lying excitations. Unfortunately, it is still 
hard to come up with a clear picture of the low-energy sector of the KAF. 
Resonating Valence Bonds (RVB) states, for which wave functions
are products of pair singlets, seem to be a natural framework to describe this
exponential proliferation of singlet states. RVB states
were first proposed to describe a disordered spin
liquid phase by Fazekas and Anderson \cite{Fazekas} for the triangular
lattice and was reintroduced by  Anderson \cite{Anderson} in the
context of high-$T_c$ superconductivity. 

For the {\it kagom\'e} lattice, the absence of long
range correlation may lead to consider only Short Range RVB states (SRRVB)
, {\em i.e.} first neighbor coverings of the lattice with dimers. 
The first difficulty which occurs is that the number of SRRVB states of a
$N$ site  {\it kagom\'e} lattice with periodic boundary conditions is
$2^{1+(N/3)}\sim 2 (1.26)^N$
\cite{Elser} whereas the number
of singlet states before the first triplet of the KAF scales like
$1.15^{N}$ \cite{Waldtmann}. Of course, this does not necessarily disqualify the SRRVB
description but raises the question of the selection of the relevant states.

\begin{figure}
\begin{center}
\resizebox{0.45\textwidth}{!}{%
  \includegraphics{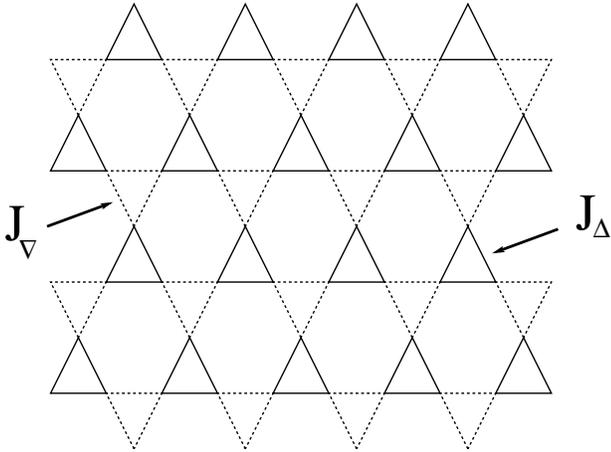}
}
\caption{Trimerized {\it kagom\'e} lattice: strong bond triangles form a
(N/3) site triangular lattice.}
\label{fig:trimerized}       
\end{center}
\end{figure}

At the mean field level an answer to this question has been given in a recent
paper \cite{Mila} starting from a trimerized \cite{remark} version of the KAF (see
Fig. \ref{fig:trimerized}): 

\begin{equation}
{\cal H}=J_{\triangledown} \sum_{\langle i,j \rangle_{\triangledown}} \vec{S}_i
. \vec{S}_j + J_{\vartriangle} \sum_{\langle i,j \rangle_{\vartriangle}}
\vec{S}_i . \vec{S}_j ,
\label{eq:trimerized}
\end{equation}

When considering low-energy excitations one can work in the subspace where
the {\em
total spin} of each strong bond triangle is $1/2$. Since there are two ways
to build a spin $1/2$ with three spins $1/2$, these triangles have two
spin $1/2$-like degrees of freedom : The total spin $\vec{\sigma}$, and the 
chirality
$\vec{\tau}$. This representation does not simplify the problem because
spin and chirality are coupled in the Hamiltonian but it is no longer the
case in the mean field
approximation and it is possible to solve the mean-field equations exactly  \cite{Mila}. Low-energy states
are SRRVB states on the triangular lattice formed by strong bond triangles and
their number grows like the number of dimer coverings of a $N/3$ site
triangular lattice, $1.15^N$, as can be shown using standard methods
\cite{Fisher,Kasteleyn}.

This result was established under the assumptions that
$J_{\triangledown}/J_{\vartriangle}$ is small (trimerized limit) and that
quantum
fluctuations can be treated at the lowest order (mean field
approximation). Therefore two questions remain open: What
happens beyond mean field approximation ? Can SRRVB state give a good
description of the energy spectrum in the isotropic limit?

To answer these questions we have studied the KAF Hamiltonian in the subspace of SRRVB
states with no simplifying approximation concerning the non orthogonality of
this basis. In this subspace the complete spectrum is obtained up to 36-site 
clusters in both trimerized and isotropic limit. 

The text is organized as follows: In the first part we study the trimerized
model and show that mean field predictions are robust with respect to quantum
fluctuations. In the trimerized limit, the low-energy spectrum splits into
bands in which the average number of dimers lying on one type of bonds is
fixed and the size of the lowest band scales as $1.15^N$.

Next we present the results obtained in the isotropic limit. Contrary to
what was suggested by previous studies \cite{Zengelser2} the singlet spectrum
obtained with SRRVB states is a continuum. Moreover the number of states below
a given total energy increases exponentially for all energy with the size of
the system considered. 

Finally, we compare the results obtained for KAF with the results obtained
using the same basis for a non-frustrated antiferromagnet, the Heisenberg
model on a square lattice, and we emphasize the ability of SRRVB states to
capture the specific low energy physics of  frustrated magnets.

Most of the results presented here contrast with the 
commonly admitted point of view that SRRVB states do not provide a good variational 
basis for this problem. In fact, SRRVB states
lead to specific numerical difficulties due to the fact that they are
not orthogonal to each other. A way to get around this difficulty is to neglect
overlap between states under a given threshold. However reasonable this
approximation may seem, it appears to modify the results significantly.
It turns out that this approximation is not necessary to perform exact numerical
simulations, even for large systems.  In order to clarify this point, 
some technical details about the method we used to implement symmetries of
the problem and achieve the calculations in this non-orthogonal basis are given in
an Appendix.
\section{The trimerized model}
\label{sec:1}
As stated in the introduction, the main question with the trimerized model
($J_{\triangledown}/J_{\vartriangle} < 1$) is
to know if the mean-field selection mechanism (pairing of strong bond
triangles) of low-lying singlet states is robust when quantum fluctuations
are taken into account.

{\it Fully trimerized limit -- Ground state}. Let us start with the limit $J_{\triangledown}/J_{\vartriangle}=0$. In this
limit the system consists of $N/3$ independent triangles and the SRRVB GS is
obtained by putting one dimer on each of these triangles. Since
$J_{\triangledown}/J_{\vartriangle}=0$ this state can be completed
to a SRRVB state ($N/2$ dimers) by putting the $N/6$ remaining dimer on the
$J_{\triangledown}$ bonds. The energy of such a state is $-(3/4)
J_{\vartriangle} (N/3)=-(N/4) J_{\vartriangle}$. In this limit the GS is
thus obtained by maximizing the number of dimers on the $J_{\vartriangle}$
bonds ($N/3$). 

\begin{figure}
\begin{center}
\resizebox{0.45\textwidth}{!}{%
  \includegraphics{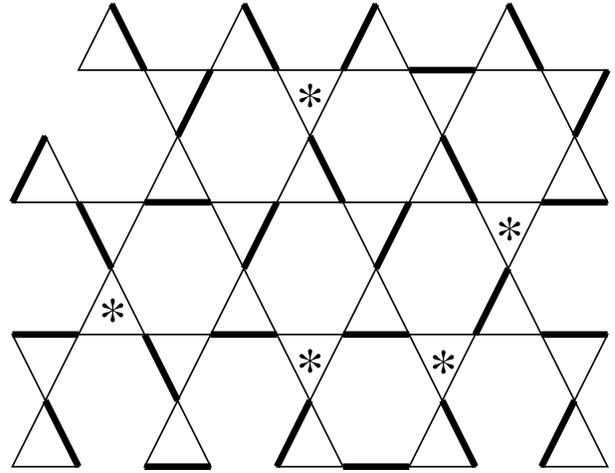}
}
\caption{Typical dimer covering of the lattice with
$n_{\text{def}}(J_{\vartriangle})=1$ (defaults are marked with a star).}
\label{fig:def}
\end{center}
\end{figure}

By a simple counting argument it is easy to see that every SRRVB state
contains $N/6=N_{t}/4$ triangles, called
defaults, for which none of the bonds is occupied by
a dimer ($N_{t}=(2N/3)$ is the number of 
triangles): a SRRVB state being a set of $N/2$ dimer, it leaves
$(2N/3)-(N/2)=(N/6)$ triangles unoccupied. The number of defaults
$n_{\text{def}}(J_{\vartriangle})$ on the
$J_{\vartriangle}$ bonds can take all the values from $0$ to $N/6$. In terms
of defaults the GS discussed above is a SRRVB state which minimizes 
$n_{\text{def}}(J_{\vartriangle})$. 

\begin{figure}
\begin{center}
\resizebox{0.45\textwidth}{!}{%
  \includegraphics{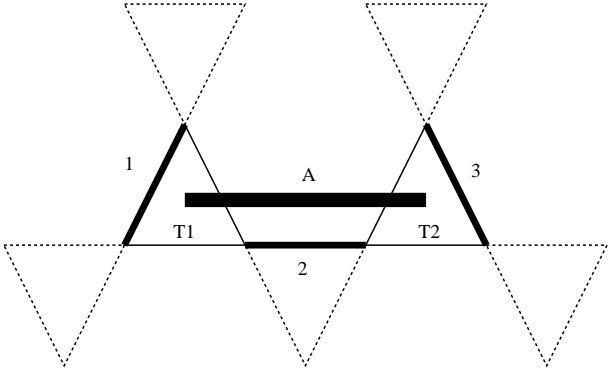}
}
\caption{The number of coverings with dimers
(A) of the $N/3$ site triangular lattice of
strong bond triangles is equal to the number of covering of the original
lattice which satisfy $n_{\text{def}}(J_{\vartriangle})=0$ (sets of $(1,2,3)$
dimer patterns).}
\label{fig:trimerized4}
\end{center}
\end{figure}

Let us turn to the question of the degeneracy of this GS and show that the number of dimer
coverings of the {\it kagom{\'e}} lattice with
$n_{\text{def}}(J_{\vartriangle})=0$ is exactly the number of dimer
coverings of the $N/3$ site triangular lattice formed by $\vartriangle$
triangles. To prove this, we have to check that one can associate each
GS configuration to a unique dimer covering of the triangular super-lattice
and vice versa (see Fig. \ref{fig:trimerized4}).

Clearly, to each pairing $A$ of  $\vartriangle$ triangles one can
associate a set of dimers $(1,2,3)$ on the {\it kagom{\'e}} lattice. Doing
so, the number of dimers on $\vartriangle$ triangles is $N/3$, which is the
maximum, and $n_{\text{def}}(J_{\vartriangle})=0$. Consider now a SRRVB with
$n_{\text{def}}(J_{\vartriangle})=0$. Let us show that there exists a unique
way to pair $\vartriangle$ triangles according to the $(1,2,3)$ pattern.
Starting from dimer $1$ on triangle $T1$, the existence of dimer $2$ is
necessary because the state is SRRVB and the triangle $T2$ contains dimer
$3$ because there is no default on $\vartriangle$ triangles by assumption.

For the triangular lattice, the number of coverings increases with the
number of sites $N$ like $A \alpha_{\text{t}}^N$ with $\alpha_{t}=\exp \{ \frac{1}{16 \pi^2} \int_0^{2\pi} \int_0^{2\pi} \ln (4+4\sin x
\sin y + 4 \sin^2 y) dx dy \} \simeq 1.5351$ and $A \simeq 2$ \cite{Mila}. Thus the
number of dimer coverings of the {\it kagom\'e} lattice with
$n_{\text{def}}(J_{\vartriangle})=0$ increases like
$(\alpha_{\text{t}}^{1/3})^N \sim 1.1536^N$. 

This degeneracy has been obtained considering only SRRVB subspace. In the
full $S=0$ subspace the GS is much more degenerate. The model, when
$J_{\triangledown}/J_{\vartriangle}=0$, simply reads:

\begin{equation}
{\cal H}=(J_{\vartriangle}/2) \sum_{i}  \left \{
S_{\vartriangle_i}(S_{\vartriangle_i}+1)-(9/4) \right \},
\label{eq:limit}
\end{equation}
where $S_{\vartriangle_i}$ is the total spin of the triangle $i$.

The GS is thus obtained by setting the total spin of each $\vartriangle$
triangle to $1/2$ and to couple all
the $N/3$ spin $1/2$ triangles to a total spin of $0$ and the degeneracy is
$2^{N/3}\;(N/3)!/[(N/6)!(1+N/6)!]$. The combinatory factor is
the size of the singlet sector of $N/3$ spin $1/2$ and the other factor
refers to the fact that on each of the $\vartriangle$ triangles there are 2
independent ways to build a total spin of $1/2$. 

Thus asymptotically
the full singlet degeneracy increases like
$2^{2N/3} / N^{3/2} \sim 1.5874^N / N^{3/2}$. The table
\ref{tab:deg} summarizes the various degeneracies.

\begin{table} [h]
\begin{center} 
\begin{tabular}{lccccc}
\hline\noalign{\smallskip}
\bf \# of sites & \bf 12 & \bf 24 & \bf 36 &  $\bf N$ \\ 
S=0 &  132 & $2.1\;10^5$ &  $4.8\;10^8$&  $\sim \frac{2^N}{N^{3/2}}$ \\
GS deg. (S=0) & 32  & 3584 & $5.4\;10^5$ &  $\sim \frac{1.5874^N}{N^{3/2}}$ \\
SRRVB & 32 &  512 & 8192 & $\sim 1.26^N$ \\ 
GS deg. (SRRVB) & 12 & 72 & 348 &  $\sim 1.1536^N$ \\ 
\noalign{\smallskip}\hline
\end{tabular}
\vskip.5cm
\caption{Number of singlet and SRRVB states and degeneracy of the GS in each
of these subspaces as a function of the number of sites.}
\label{tab:deg}
\end{center}
\end{table}

\begin{figure}
\begin{center}
\resizebox{0.45\textwidth}{!}{%
  \includegraphics{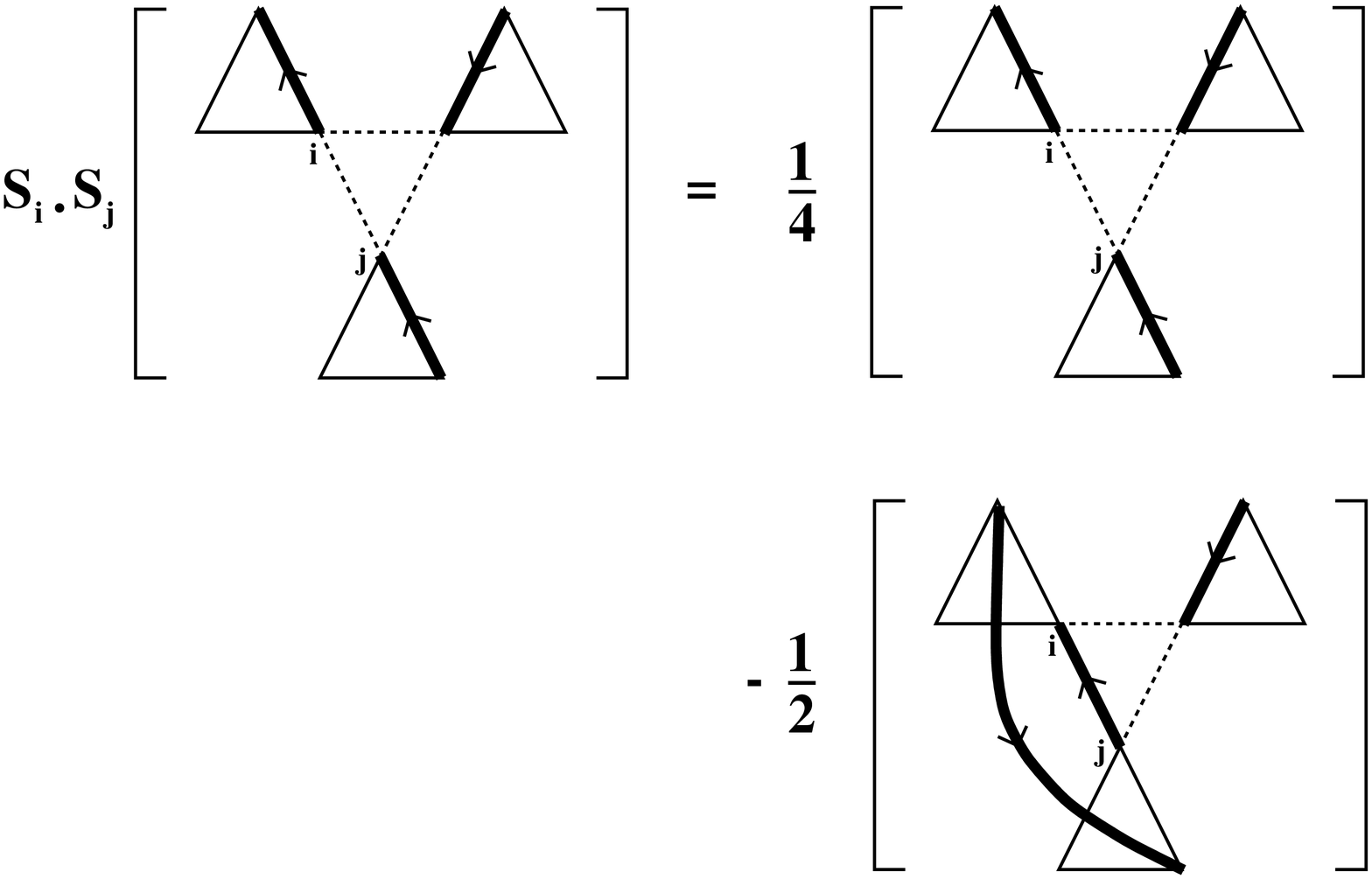}
}
\caption{In the most general case, when a covering includes at least one
non-zero bond triangle default, SRRVB states are not
eigenstates of $\cal H$: off-diagonal terms, which overlap with all SRRVB
coverings, are generated.}
\label{fig:trimerized5}
\end{center}
\end{figure}

{\it Fully trimerized limit -- Excited states}. The situation of excited
states in the SRRVB subspace is
different, even when $J_{\triangledown}/J_{\vartriangle}=0$, because SRRVB
states with $n_{\text{def}}(J_{\vartriangle}) \neq 0$ are not eigenvectors
of ${\cal H}$. In fact this situation occurs each time a state includes a
default on a triangle with non-zero bonds (see Fig. \ref{fig:trimerized5}).

\begin{figure}
\begin{center}
\resizebox{0.45\textwidth}{!}{%
  \includegraphics{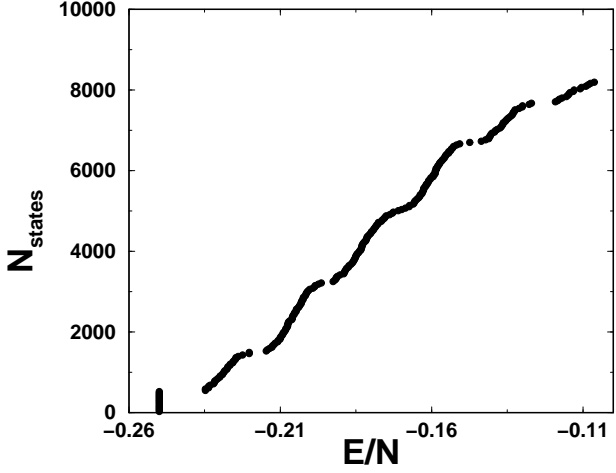}
}
\caption{Number of states (y-axis) with an energy per site smaller than
$E/N$ (x-axis) for
a 36 site cluster with $J_{\triangledown}/J_{\vartriangle} = 0$. The
spectrum splits into bands in which $n_{\text{def}} (J_{\vartriangle})$ is
fixed.}
\label{fig:trimerized0}
\end{center}
\end{figure}

Nevertheless, let us consider the results obtained for
$J_{\triangledown}/J_{\vartriangle}=0$ (Fig. \ref{fig:trimerized0}). 
The spectrum splits into bands: the first, of
zero width, is the degenerate GS discussed above, and the other bands consist of linear
combinations of SRRVB states with fixed $n_{\text{def}}
(J_\vartriangle)$ (a numerical characterization of the dimer coverings in
each band is given below). The center of each of these bands is $-(3/4)
N_{\vartriangle} J_{\vartriangle}$ with $N_{\vartriangle}$ the number of
dimers built on $\vartriangle$ triangles. Since
$N_{\vartriangle}=(N/3)-n_{\text{def}} (J_{\vartriangle})$, the energy of the
center of the $1+(N/6)$ bands are  $-(N/4)
J_{\vartriangle}$,$((3/4)-(N/4)) J_{\vartriangle}$,\ldots,$-(N/8)
J_{\vartriangle}$.

\begin{figure}
\begin{center}
\resizebox{0.45\textwidth}{!}{%
  \includegraphics{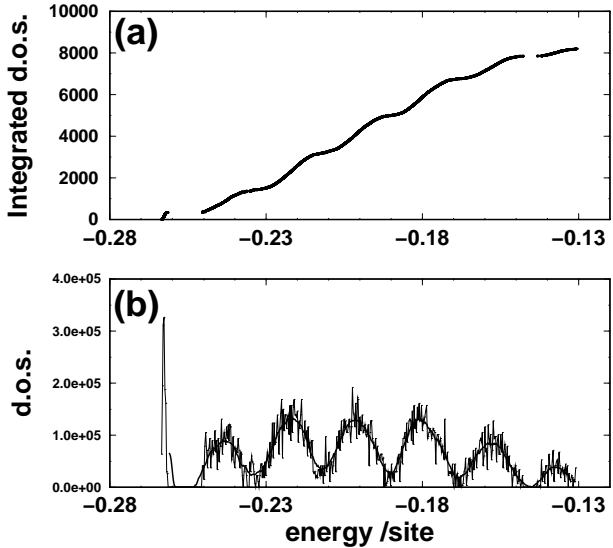}
}
\caption{(a) Same as Fig. \ref{fig:trimerized0} for a 36 site cluster with
$J_{\triangledown}/J_{\vartriangle} = 0.1$. (b) Density of state (derivative
of (a)). The band structure of the spectrum is conserved for a weak
trimerization. While the excited states band begin to mix the GS band belongs
separated from the others.}
\label{fig:trimerized1}
\end{center}
\end{figure}

{\it Strong trimerization} ($J_{\triangledown}/J_{\vartriangle} \ll 1$). When it is switched on, $J_{\triangledown}$ acts as a perturbation on the
previous spectrum: bands with $n_{\text{def}} (J_{\triangledown}) \neq 0$
begins to get wider and to mix. In contrast, because it is degenerate when
$J_{\triangledown} / J_{\vartriangle}=0$, the lowest band is expected to
mix with the other bands for larger values of
$J_{\triangledown}/J_{\vartriangle}$. Let us test this scenario on numerical
results for a weak trimerization of the
lattice ($J_{\triangledown}/J_{\vartriangle}=0.1$) on a 36 site cluster.
Figure \ref{fig:trimerized1} shows the spectrum (number of states below a
given energy per site) and the density of states (DOS) . The DOS exhibit a
band structure and, as expected, even for such a small value of
$J_{\triangledown} / J_{\vartriangle}$, 
gaps between bands are nearly closed. Nevertheless, a very narrow band of
states remains very clearly separated from the others. 

The existence of this low energy band splited from the rest of the SRRVB
spectrum indicates that for small values of
$J_{\triangledown}/J_{\vartriangle}$, the selection criterion of dimer covering
configurations is the same as for $J_{\triangledown}/J_{\vartriangle}=0$:
the states in the low energy part of the spectrum minimize $n_{\text{def}}
(J_{\vartriangle})$. In order to test more precisely this scenario let
us characterize numerically the scaling of the bands and verify that
$n_{\text{def}} (J_{\vartriangle})$ is fixed in each band.

We performed a finite size analysis including all
{\it kagom\'e} clusters with an even number of sites up to 36 sites (12, 18,
24, 30 , 36). We denote by
${\cal N}_{N}(\Delta)$ the number of states on a N-site cluster with a
{\em total energy} smaller than $\Delta$. For all $\Delta$, the analysis
shows that ${\cal N}_{N}(\Delta)$ grows exponentially with $N$ :

\begin{equation}
{\cal N}_{N}(\Delta) = A(\Delta) {\alpha(\Delta)}^{N}.
\label{eq:alpha}
\end{equation}

\begin{figure}
\begin{center}
\resizebox{0.45\textwidth}{!}{%
  \includegraphics{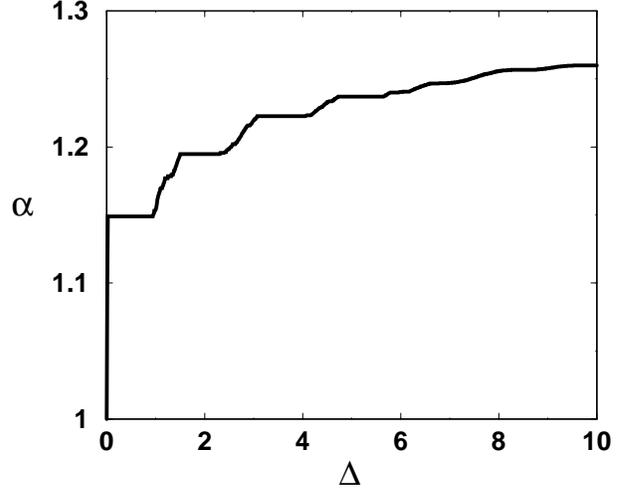}
}
\caption{Extrapolation of ${\alpha(\Delta)}$ for
$J_{\triangledown}/J_{\vartriangle} = 0.1$ from $12,18,24,30,36$ site
spectra.  ${\alpha(\Delta)}$ exhibit a plateau $(1.15)$ which corresponds to
the scaling of the lowest band of the spectrum.}
\label{fig:trimerized2}
\end{center}
\end{figure}

In the large $\Delta$ limit, since all the states have an energy smaller
than $\Delta$, the values of $A$ and $\alpha$ are known to be respectively
$2$ and $2^{1/3}$. Between each band of the spectrum, no state comes to
increase ${\cal N}$ when $\Delta$ increases and therefore plateaus appear in
$\alpha$ (see Fig. \ref{fig:trimerized2}). The first plateau corresponds to
$\alpha \simeq 1.15$, a numerical confirmation of what was announced
at the beginning of the section.

\begin{figure}
\begin{center}
\resizebox{0.45\textwidth}{!}{%
  \includegraphics{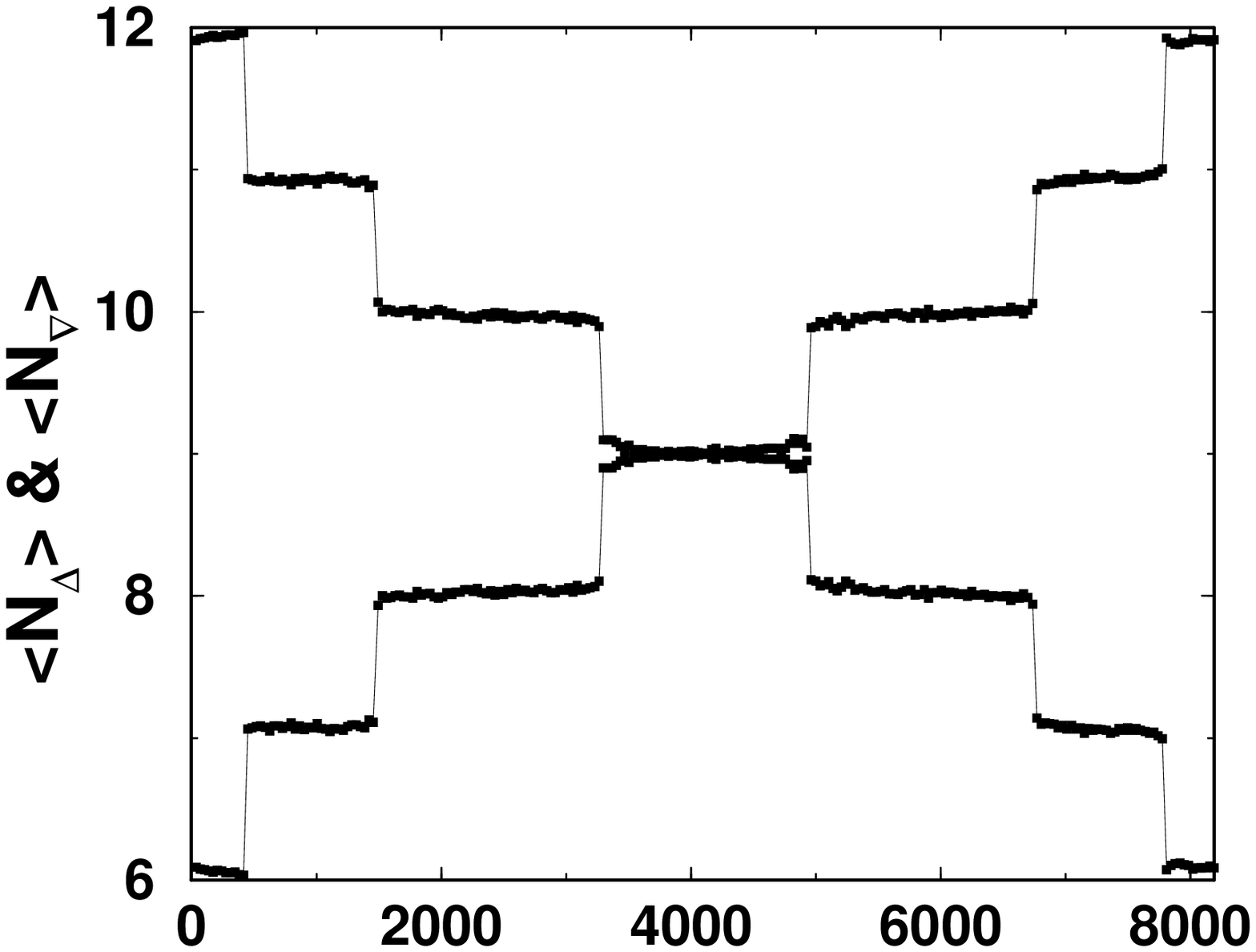}
}
\caption{Average values of $\hat{N}_{\vartriangle}$ and
$\hat{N}_{\triangledown}$ (y-axis) for each eigenstate of
$\cal H$ (from GS to most excited state, x-axis) for a 36 site cluster with $J_{\triangledown}/J_{\vartriangle} = 0.1$.}
\label{fig:trimerized3}
\end{center}
\end{figure}

Let us turn now to the question of the nature of the states in
each band. We denote by $\hat{N}_{\triangledown}$ and
$\hat{N}_{\vartriangle}$ the operators that count for a SRRVB state the
number of dimers lying on $J_{\triangledown}$ and $J_{\vartriangle}$ bonds
respectively. Since on a $N$ site cluster, each SRRVB state is made of $N/2$
dimers, we have $\hat{N}_{\triangledown}+\hat{N}_{\vartriangle}=N/2$.
Fig. \ref{fig:trimerized3} shows the values of $\langle
\hat{N}_{\triangledown} \rangle$ and $\langle \hat{N}_{\vartriangle}
\rangle$ for each eigenstate of a 36 site cluster from the GS to the most
excited state. The results are quite
clear: each band of the spectrum is characterized by a fixed value of
$\langle \hat{N}_{\vartriangle} \rangle$ (or $\langle
\hat{N}_{\triangledown} \rangle$) which is equivalent to fix $\langle
n_{\text{def}} (J_\vartriangle) \rangle = (N/3) - \langle
\hat{N}_{\vartriangle} \rangle$.

{\it SRRVB spectrum versus Exact spectrum}. 
SRRVB states on the trimerized {\it kagom\'e} lattice spontaneously selects
a small set of wave functions (see table \ref{tab:deg}) among those which
minimize energy for $J_{\triangledown}/J_{\vartriangle}=0$.
Moreover the number of these
states scale as the number of singlets in the singlet-triplet gap of the KAF
at the isotropic limit. If this selection is actually relevant, one should be
able to identify  in the {\it exact spectrum} at least for a strong
trimerization the existence of a similar selection.

To test this point we compare the exact and SRRVB spectra for
$J_{\triangledown}/J_{\vartriangle}=0.25$ (see table \ref{tab:spect},  energy per site for the 10 first states). The conclusion of this comparison  is quite clear: The SRRVB
subspace reproduces the low-energy part of the singlet spectrum and the
structure of the spectrum (order and degeneracy of levels) is also well
described.

\begin{table} [h]
\begin{center}
\begin{tabular}{ccc}
\hline\noalign{\smallskip}
\bf Exact Diag. & \bf SRRVB & \bf Deg.  \\
-0.29530 & -0.29513 & 1 \\ 
-0.29049 & -0.28714 & 2  \\ 
-0.29027 & -0.28644 & 1 \\ 
-0.28597 & -0.28453 & 3 \\ 
-0.28187 & -0.28053 & 3  \\
\noalign{\smallskip}\hline
\end{tabular}
\vskip.5cm
\caption{Comparison between exact and SRRVB low energy levels for a 12 site
cluster with $J_{\triangledown}/J_{\vartriangle}=0.25$.}
\label{tab:spect}
\end{center}
\end{table}

In conclusion, beyond mean field approximation, the low energy physics of
the trimerized KAF is well captured by SRRVB states: Low lying states are
selected on an energy criterion, the maximization of the number of dimers on strong
bonds, which is equivalent for a weak trimerization to minimize the number of
defaults on strong-bond triangles. These selected
states form a band which contains a number of states that increases like $1.15^N$ in agreement both with
exact results and with the effective Hamiltonian approach. 

\section{The isotropic model}
\label{sec:2}

When $J_{\triangledown}/J_{\vartriangle}$ increases to the isotropic limit
one may ask at least two questions: Does the mechanism described above
remain valid? Do SRRVB states still provide a good description of the
singlet sector?

\begin{figure}
\begin{center}
\resizebox{0.45\textwidth}{!}{%
  \includegraphics{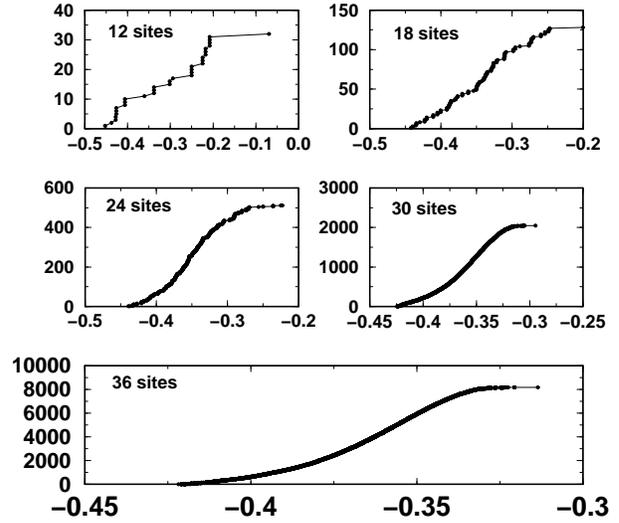}
}
\caption{Integrated d.o.s of 12,18,24,30,36 site clusters at the
isotropic point. The spectrum has the structure of a continuum.}
\label{fig:ntrimerized1}
\end{center}
\end{figure}

To answer the first question, we have computed, at the isotropic point, the values
of $\langle \hat{N}_{\triangledown} \rangle $ and $\langle
\hat{N}_{\vartriangle} \rangle$ for all the eigenstates. The behavior of
these quantities is very different from the trimerized case: no mechanism
tends to favor one type of triangle and $\langle \hat{N}_{\triangledown}
\rangle = \langle \hat{N}_{\vartriangle} \rangle = N/4$. This means that the
simple picture obtained with the trimerized model is no longer valid in the
isotropic case. The computation of the spectrum for all even sizes up to 36
sites confirms the qualitative differences between trimerized and isotropic
model (see Fig. \ref{fig:ntrimerized1}). The mixing of the bands which
starts for $J_{\triangledown} < J_{\vartriangle}$ is complete for
$J_{\triangledown} = J_{\vartriangle}$, the band structure has completely
disappeared and the spectrum is a {\em continuum}.

It is important to emphasize that this result contrasts with
those obtained for the same model by Zeng and Elser\cite{Zengelser2}, 
who
concluded to the presence of a gap inside the singlet spectrum. This study
was based on an expansion using as a small parameter the overlap between SRRVB
states: the non orthogonality between dimer covering $\vert \varphi_i
\rangle $ was neglected under a given threshold of $\langle \varphi_i \vert
\varphi_j \rangle$. On the contrary, the results presented
here involve no approximation: the non orthogonality of the basis is fully
taken into account (see appendix for details). We suspect that the
difference comes from this approximation. As could be expected, our
treatment provides a smaller variational value of the GS energy. For a 36
site cluster $E_{\text{GS}}/J=-0.42182$  which is $3\%$
above the exact one (the number of SRRVB states is $\sim 1.71\;10^{-5}$ 
of the total singlet subspace). 

\begin{figure}
\begin{center}
\resizebox{0.40\textwidth}{!}{%
  \includegraphics{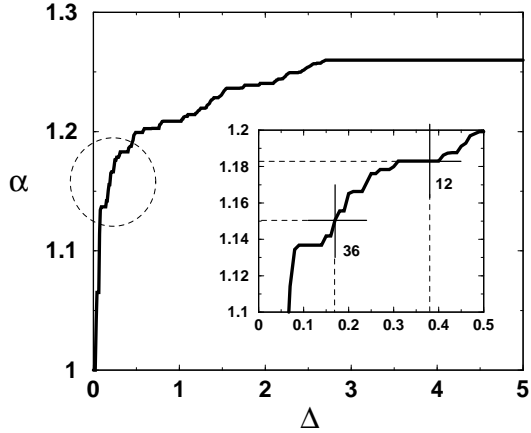}
}
\caption{$\alpha(\Delta)$ (see Eq.\ref{eq:alpha}) for the standard {\it
kagom\'e} model. {\it Inset:} zoom of the circled region. The abscissa of the
points denoted by the crosses are the {\it exact} diagonalization
values of the singlet-triplet
gap for the 36 site and 12 site clusters.}
\label{fig:ntrimerized2}
\end{center}
\end{figure}

The strongest indication that SRRVB states give a correct
description of the low-lying singlets of the KAF
is indeed the continuum structure of the spectrum. Moreover, the shape of
the spectrum is very similar to the one obtained by exact diagonalization (ED)
\cite{Waldtmann}. 
In order to test this point more precisely, we have again computed
$\alpha(\Delta)$ at the isotropic point (see Fig. \ref{fig:ntrimerized2}). Plateaus no longer appear in
$\alpha(\Delta)$, which confirms the complete mixing of the bands. More
interestingly, this analysis shows that, {\em for all $\Delta$, the number of
SRRVB excitations increases exponentially with the size of the systems}.
This proves that SRRVB states not
only reproduce the continuum nature of the spectrum but give a good
description of the exponential proliferation of singlets states in the low
energy sector of the KAF.

Since the SRRVB subspace cannot give information about magnetic excitations
the question of the counting of states below the first triplet is
rather delicate. To discuss this point, one has to take the {\it exact}
singlet-triplet gap value to make the counting in the {\it variational} SRRVB
spectrum. Doing so on has to keep in mind that even if the SRRVB spectrum
gives a good description of the structure of the low lying singlets (order of
levels, degeneracy, exponential proliferation) the energy scale of the
excitations above the GS might be slightly different from the exact one:
SRRVB are not the exact eigenstates of the Hamiltonian which are more
probably dressed SRRVB states including fluctuations that modify the energy
scale. But, given the relative accuracy of the GS this point should not
prevent us from doing a semi-quantitative comparison between exact and SRRVB
results. 

In fact, for a 12 site cluster we have checked that the low energy structure
of excitation spectrum is correct for $J_{\triangledown}/J_{\vartriangle}=1$
up to the first triplet state (see Table \ref{tab:spect2}).

\begin{table} [h]
\begin{center}
\begin{tabular}{cccc}
\hline\noalign{\smallskip}
\bf Exact Diag. & \bf SRRVB & \bf Deg.  & \bf S\\ 
-0.45374 &  -0.45313 & 1 &  0\\ 
-0.44403 &  -0.43764 & 1 &  0\\ 
-0.44152 &  -0.42803 & 2 &  0\\
-0.43044 &  -0.42703 & 3 &  0\\ 
-0.42185 &  XXX & 9 &  1 \\ 
-0.41438 &  -0.40625 & 3 &  0\\ 
\noalign{\smallskip}\hline
\end{tabular}
\vskip.5cm
\caption{Comparison between exact and SRRVB low energy levels for a 12 site
cluster at the isotropic point $J_{\triangledown}/J_{\vartriangle}=1$.}
\label{tab:spect2}
\end{center}
\end{table}

For a more general quantitative discussion on the proliferation of
low-energy singlets let us analyze the shape of $\alpha(\Delta)$ (see
Fig. \ref{fig:ntrimerized2}) obtained from SRRVB spectra of 12, 18, 24, 30
and 36
site clusters. The range of the {\it exact} singlet-triplet gap extends
from $\sim 0.38 J$ for the 12 site cluster to $\sim 0.17 J$ for 36 sites
\cite{Waldtmann} which corresponds to the circled region and the inset of
Fig. \ref{fig:ntrimerized2}. It is remarkable that in this energy range the
value of $\alpha$ for SRRVB spectra goes from  $\sim 1.18$ to $\sim 1.15$
which in good agreement with ED scalings.

\begin{figure}
\begin{center}
\resizebox{0.45\textwidth}{!}{%
  \includegraphics{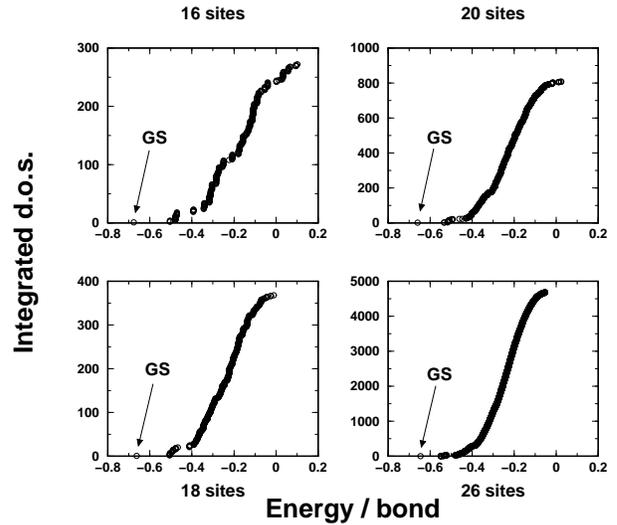}
}
\caption{Integrated d.o.s. for the Heisenberg antiferromagnet on the square
lattice in the SRRVB subspace for 16, 18, 20 and 26 site clusters. The low
energy structure of the spectrum is no longer a continuum.}
\label{fig:carre}
\end{center}
\end{figure}

\section{Discussion}
\label{sec:3}

At this point, it is fair to ask whether the continuum structure of the
spectrum obtained with SRRVB states is really a specific feature of
frustration captured by this basis or simply a generic characteristic
of the spectra that such states would provide on any lattice. To answer
this important question let us compare the results on the {\it kagom\'e}
lattice with the SRRVB spectrum for a
non-frustrated model, the Heisenberg model on the square lattice (see
Fig. \ref{fig:carre}).

\begin{figure}
\begin{center}
\resizebox{0.45\textwidth}{!}{%
  \includegraphics{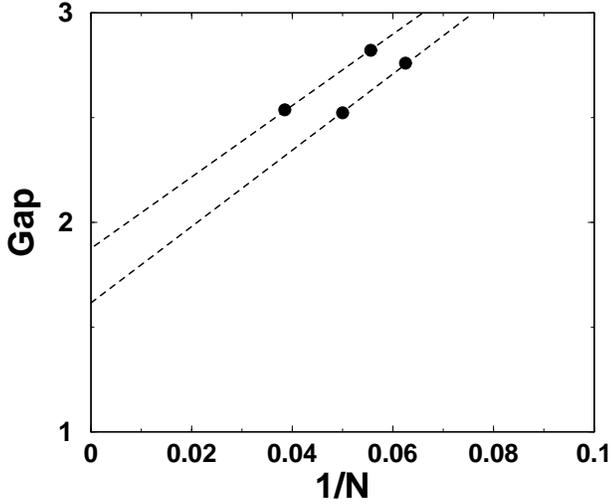}
}
\caption{Total energy difference between SRRVB first
excited state and GS for the square lattice (16, 18, 20 and 26 site
clusters). The Data suggest a non-zero value for the gap in the  thermodynamic limit.}
\label{fig:carregap}
\end{center}
\end{figure}

The structure of the SRRVB excitations on the square lattice is
qualitatively different from the structure obtained for the {\it kagom\'e}
lattice: In particular there is a gap
between the singlet GS and the first excitation. Even if it is seems
difficult to extract a precise value for this gap (see Fig.
\ref{fig:carregap}), the finite size analysis strongly suggests that it
remains finite in the thermodynamic limit. Of course, this does {\it not} describe 
the actual singlet spectrum of the square lattice, which is gapless due to 
two-magnon excitations. But it shows that the structure of the RVB spectrum is 
specific to very frustrated lattices.

In conclusion, SRRVB states on the {\it kagom\'e} lattice allow to capture the
specific low energy properties of the model in both trimerized and isotropic
limits. In the trimerized model it gives a simple picture of the non magnetic
excitations and a selection criterion of the low-energy states which are
built by minimizing the number of defaults on strong bond triangles. The
number of such states increases like $1.15^N$. The states matching this
criterion can also be seen as short-range dimer coverings of the triangular
lattice formed by strong-bond triangles which confirms, beyond mean field
approximation, the relevance of the effective model approach.
At the isotropic point, SRRVB states lead to a continuum of
non-magnetic excitations in accordance with ED results. Moreover the shape
of the SRRVB spectrum is very similar to the exact one and the number of
low-lying singlets increases exponentially for all energy range with the
size of the system considered. 

Finally, these properties of the SRRVB spectrum are {\em not} just a
property of this kind of states since the SRRVB spectrum has a gap in the
case of the square lattice. So one may conjecture that they provide a good
description of the low-energy singlet sector of very frustrated magnets
only. Work is in progress to test this idea on the pyrochlore lattice.

{\it Acknowledgments: } We acknowledge useful discussions with
C.~Lhuillier, B. Dou\c{c}ot and P.~Simon. We are especially grateful to
P.~Sindzingre for making available unpublished results of exact
diagonalization on the {\it kagom\'e} lattice.

\section{Appendix: numerical method}
\label{sec:4}

Working with SRRVB states as a truncated basis leads
to non-trivial numerical difficulties which, as paradoxical as it may seem,
make the problem of the determination of the spectrum more tricky in this
truncated subspace than in the full space of spin
configurations. This is a consequence of the non-orthogonality of
RVB states.

\begin{figure}
\begin{center}
\resizebox{0.35\textwidth}{!}{%
  \includegraphics{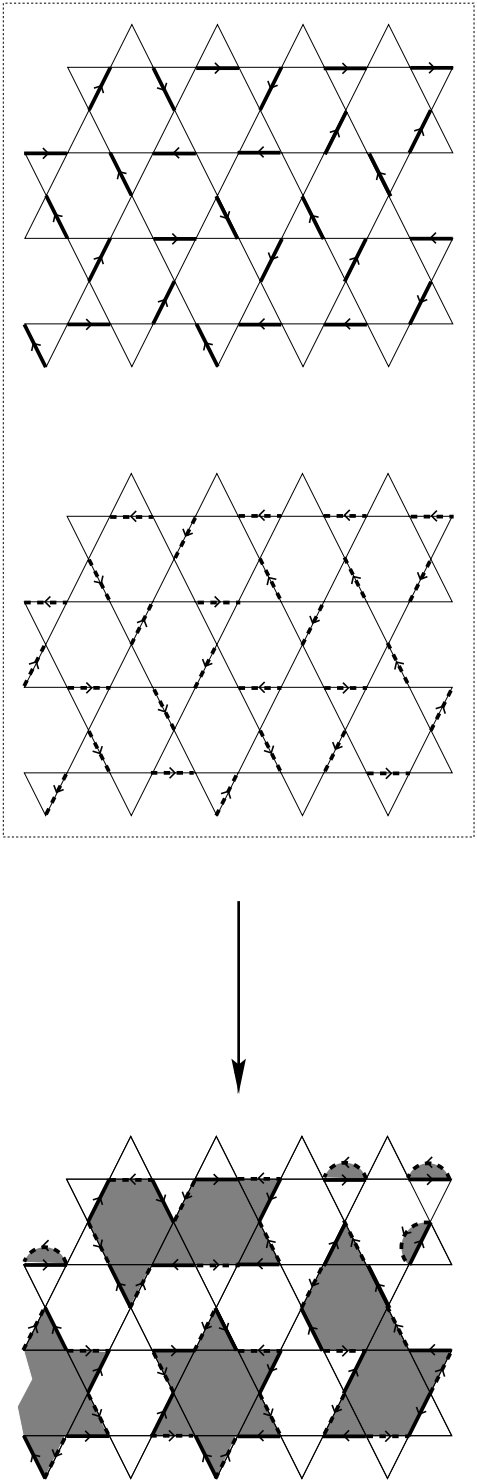}
}
\caption{The overlap between two dimer coverings depends on the number of
loops in the transition graph (bottom). Note that the relative orientation
of dimer coverings in this figure corresponds to the reference orientation
which produces a positive overlap.}
\label{fig:ann1}
\end{center}
\end{figure}

If $\vert \varphi_i \rangle$ and $\vert \varphi_j \rangle$ are
two SRRVB states, the overlap is given by \cite{Sutherland}

\begin{equation}
\langle \varphi_i \vert \varphi_j \rangle = s(\varphi_i,\varphi_j) \cdot 2^{n_{b}(\varphi_i,\varphi_j)-N/2},
\label{eq:overlap}
\end{equation}
where $n_{b}(\varphi_i,\varphi_j)$ is the number of closed loops in the
diagram where the two states are superimposed and
$s(\varphi_i,\varphi_j)=(-1)^p$,  where $p$ is the number of misoriented dimers
compared with the reference orientation (see Fig. \ref{fig:ann1}).

In the case of a non-orthogonal basis, the eigenvalues are 
solutions of the so-called generalized eigenvalue problem,

\begin{equation}
\det ( \langle
\varphi_i \vert {\cal H} \vert \varphi_j \rangle - E \langle \varphi_i \vert
\varphi_j \rangle )=0,
\label{eq:ann2}
\end{equation}
in which the overlap between states appears
explicitly. Since we are interested in the structure of the spectrum, we need to
diagonalize completely the Hamiltonian and therefore iterative techniques
(typically Lanczos) must be avoided. On the other hand, solving
(\ref{eq:ann2}) with standard routines, one is limited to small systems.
  
To achieve a complete diagonalization for large systems (typically 36 sites) it
is crucial to take into account all the symmetries of the system in order to
break the Hilbert space into smaller pieces. This technique is indeed very
standard but is usually used in a context where the basis is {\it
orthogonal} (e.g. spin configurations) which makes it quite convenient. The
non-orthogonal case is far less simple and is worth paying some attention.

The aim of this appendix is to explain how it is
possible, starting from a set of configurations that can be non-orthogonal,
to build an {\it orthonormal} basis of vectors that are {\it eigenstates} of
all the symmetries of the problem in each symmetry sector. Since this linear
algebra problem is planned to be solved numerically one is interested in
reducing as much as possible the information to be handled. Therefore one
does not work explicitly with this orthonormal basis but with linear
combinations of suitably chosen configurations called representatives.

The text is organized as follow: we define the representatives, we show how the
number of representatives has to be reduced depending on the symmetry sector
and finally explain how one can go from representatives to the orthogonal
basis of the symmetries eigenvectors.

{\it Representatives.}  Let us denote by $N_{\cal S}$ the order of the symmetry group of the system and
${{\cal S}_{i}},\;i=1,\dots,N_{\cal S}$ the elements of this group. It is
possible to make a partition of the set containing all the configurations in
subsets where configurations are related to each other by a symmetry ${\cal
S}_{i}$. Each of these $N_p$ subset can be represented by a
configuration $\vert p_i\rangle,\;i=1,\dots,N_{p}$, called  representative,
of the subset since by construction all the others can be obtained by
applying symmetries on it. From a numerical point of view the set of the
representatives is the minimal information needed. 

{\it Reduction of the  number of representatives in a given symmetry sector.}
In this section we will consider a given symmetry sector $s$ characterized
by a set of characters $\chi_{s} ({{\cal S}}_{1})$,\ldots,$\chi_{s} ({{\cal
S}}_{N_{\cal S}})$. We are going to show that it is not necessary to keep
all the representatives to generate a basis in the sector $s$.

Let us consider
a given linear configuration of representatives,
\begin{equation}
\vert \psi \rangle = \sum_{i=1}^{N_{p}} \alpha_i \; \vert p_i\rangle.
\label{eq:ann3}
\end{equation}
The true state of $s$ associated to $\vert \psi \rangle$ is given by \cite{remark1},
\begin{equation}
\vert \tilde{\psi} \rangle \triangleq\frac{1}{\sqrt{N_{\cal S}}} \sum_{i=1}^{N_{\cal S}} {{\cal
S}_{i}} \; \vert \psi \rangle=\sum_{i=1}^{N_{p}} \sum_{p=1}^{N_{\cal S}} 
\alpha_i \; \chi_{s} ({{\cal S}}_{p}) \; \vert s_{p}(p_i)\rangle,
\label{eq:ann4}
\end{equation}
where $\vert s_{p}(p_i)\rangle$ stands for the image of the configuration
$p_i$ by the symmetry ${{\cal S}}_{p}$. For a given representative $\vert
p_i\rangle$, let us denote by ${\cal E}_i$ the set of indices $q$ of the
symmetries that leave 
the configuration $p_i$ invariant ($s_{q}(p_i)=p_i$), and $\bar{{\cal E}_i}$
the remaining indices. With this notation $\vert \tilde{\psi} \rangle$
take the form,

\begin{eqnarray}
\vert \tilde{\psi} \rangle &  = & \sum_{i=1}^{N_{p}}
\alpha_i \left ( \sum_{\substack{q \in {\cal E}_i}}
\chi_{s} ({{\cal S}}_{q}) \right ) \vert p_i\rangle \nonumber \\
 && +  \sum_{i=1}^{N_{p}}
\sum_{\substack{q \in \bar{{\cal E}_i}}}
\alpha_i \; \chi_{s} ({{\cal S}}_{q}) \; \vert s_{q}(p_i)\rangle.
\label{eq:ann3c}
\end{eqnarray}

Let us  denote by ${\cal Q}_s$ the list of indices $i$ of the
representatives $p_i$ such as $\sum_{\substack{q \in {\cal E}_i}}
\chi_{s} ({{\cal S}}_{q}) =0$ in the symmetry sector $s$. It is obvious
to note that all the representatives with an index in ${\cal Q}_s$ disappear
from the first term of eq.~(\ref{eq:ann3c}). What we are going to show is
that they also disappear from the second one. Let be $p_i$ such as
$\sum_{\substack{q \in {\cal E}_i}} \chi_{s} ({{\cal S}}_{q}) =0$
and $p \in {\cal E}_i$. One has $\sum_{\substack{q \in \bar{{\cal E}_i}}} {{\cal
S}}_{q} \vert p_i\rangle= \sum_{\substack{q \in \bar{{\cal E}_i}}} \chi_{s}^{*}
({{\cal S}}_{p}) {{\cal
S}}_{p} {{\cal S}}_{q} \vert p_i\rangle= \sum_{\substack{q \in \bar{{\cal E}_i}}} \chi_{s}^{*}
({{\cal S}}_{p})  {{\cal S}}_{q} \vert p_i\rangle$. This result does not
depend on  $p \in {\cal E}_i$ and thus one does not modify the result by
applying $[1/{\text{Card}} ({{\cal
E}_i})] \sum_{\substack{p \in {\cal E}_i}}$ on the previous expression which proves what was announced. This 
leads to a reduction of
the number of representatives in the symmetry sector $s$, namely
$N_s=N_p-{\text{Card}} ({\cal Q}_s)$.

{\it From non-orthogonal representatives to the orthonormal basis of symmetries
eigenvectors.} In the general case of non-orthogonal representatives, it is
convenient to introduce a mixing matrix $\mu^{s}$ in order to build the
orthonormal basis of symmetries eigenvectors. We will considerer from now
linear combinations of mixed representatives,
\begin{equation}
\vert l \rangle = \sum_{j=1}^{N_{s}} \mu^{s}_{lj} \vert
p_j\rangle.
\label{eq:ann3a}
\end{equation}

All the problem is to chose $\mu^{s}$ such as the symmetrized states $\vert \tilde{l}
\rangle$ of $\vert l \rangle$ according to (\ref{eq:ann4}) form an {\it
orthonormal} basis: $\langle \tilde{l} \vert \tilde{m}
\rangle=\delta_{lm}$. Let us show how this condition writes in the cases
of orthogonal and non-orthogonal representatives.

{\it Orthogonal case}. In this simple case where $\langle p_i \vert p_j \rangle=\delta_{ij}$, the condition $\langle \tilde{l} \vert \tilde{m}
\rangle=\delta_{lm}$ is,

\begin{align}
\delta_{lm} &=
\sum_{i,j=1}^{N_s} \sum_{q=1}^{N_{\cal
S}}  {\mu^{s*}_{jl}} \mu^{s}_{mi} \langle p_j \vert {{\cal
S}}_{q} \vert p_i\rangle \nonumber \\
&= 
\sum_{i,j=1}^{N_s} \mu^{s*}_{jl}  \mu^{s}_{mi} \left (
\sum_{q \in {\cal E}_i}
\chi_{s} ({{\cal S}}_{q}) \right ) \delta_{ij} \nonumber \\
&= 
\sum_{i,j=1}^{N_s} \mu^{s*}_{jl}  \mu^{s}_{mi} {\text{deg}}(p_i)\delta_{ij},
\label{eq:ann5}
\end{align}

where $\chi_{s} ({{\cal S}}_{p})$ is the character of the symmetry ${{\cal
S}}_{p}$ in the sector $s$, ${\text{deg}}(p_i)$ the degeneracy of
representative $p_i$ (i.e. the number of symmetries under which it is
invariant \cite{remark2}), $s_{p}(p_i)$ the image of the configuration 
$p_i$ by ${{\cal
S}}_{p}$, and $N_s$ the size of the sector $s$.

It is easy
to see that 
\begin{equation}
\mu^{s}_{ij}=\delta_{ij} / \sqrt{{\text{deg}}(p_i)}
\label{eq:ann6}
\end{equation}
fulfills (\ref{eq:ann5})

{\it General case}. In the non-orthogonal case, the condition (\ref{eq:ann5}) now reads,
\begin{equation}
\delta_{lm} = \sum_{i,j=1}^{N_s} \mu^{s*}_{jl}  \mu^{s}_{mi} \tilde{I}_{ij}
\label{eq:ann7}
\end{equation}
where,

\begin{equation}
\tilde{I}_{ij}= \sum_{p=1}^{N_{\cal S}} \chi_{s} ({{\cal S}}_{p}) \langle
p_j \vert s_{p}(p_i)\rangle
\label{eq:ann8}
\end{equation}

Here again, the indices $i$ and $j$ runs from $1$ to $N_s$ (the size of the
sector $s$). To determine $\mu^{s}$ we diagonalize $\tilde{I}$ : 
\begin{equation}
P^{\dagger} \tilde{I} P = {\text{Diag}} (d_1,\dots,d_{N_s})
\label{eq:ann9}
\end{equation}

One can check that,
\begin{equation}
\mu^{s}_{ij}=\frac{1}{\sqrt{d_i}}\;P_{ij}
\label{eq:ann10}
\end{equation}
satisfies condition (\ref{eq:ann7}).

The basis $\{ \vert \tilde{l} \rangle \}$ is orthogonal and in this new
basis the Hamiltonian is block diagonal, each block corresponding to one
symmetry sector. Thus, it only remains to diagonalize the Hamiltonian in
each of these representations to get the whole spectrum :

\begin{equation}
\langle \tilde{l} \vert {\cal H} \vert \tilde{m} \rangle = \sum_{i,j=1}^{N_s}  \sum_{p=1}^{N_{\cal
S}} \mu^{s*}_{jl}  \mu^{s}_{mi} \chi_{s} ({{\cal S}}_{p}) \langle
p_j \vert {\cal H} \vert s_{p}(p_i)\rangle
\label{eq:ann11}
\end{equation}

In conclusion the procedure described above turns the generalized eigenvalue
problem of $n \times n$ matrices into $2\times N_s$ conventional
diagonalizations of $\sim (n/N_s) \times (n/N_s)$ matrices. The point we now want to stress is that the treatment described above is
exact and does not introduce approximation. The subspace of RVB states is a
truncated subspace in the sense that it is not stable with respect to an
application of the Hamiltonian and only the RVB restricted Hamiltonian is
studied. But, the use of symmetries does not act as a new restriction of the
Hamiltonian in each representation of the symmetry group. In the new basis
the Hamiltonian {\em and} the overlap matrix are actually block diagonal, each block corresponding to one
representation. Thus the spectrum obtained as well as mean values of
operators calculated are the same as those one would obtain by solving
with brute-force the generalized eigenvalue problem if it was possible.


\begin{thebibliography}{}
\bibitem{Majumdar} C.K.~Majumdar and D.~Ghosh, J. Math. Phys. {\bf 10},
(1969), 1388.

\bibitem{Waldtmann} C.~Waldtmann, H.-U.~Everts, B.~Bernu, 
C.~Lhuillier, P.~Sindzingre, P.~Lecheminant, L.~Pierre, European Physical
Journal B {\bf 2}, (1998), 501.

\bibitem{Canalspriv} B.~Canals, private communication.

\bibitem{Mambrini} M.~Mambrini, J.~Tr\'ebosc and F.~Mila, Phys. Rev. B {\bf
59},(1999), 13806.

\bibitem{Zengelser1}
C.~Zeng and V.~Elser,
\newblock Phys. Rev. B \textbf{42}, (1990), 8436.

\bibitem{Singhhuse}
R.~Singh and D.~Huse,
\newblock Phys. Rev. Lett. \textbf{68}, (1992), 1766.

\bibitem{Leungelser}
P.~Leung and V.~Elser,
\newblock Phys. Rev. B \textbf{47}, (1993), 5459.

\bibitem{Zengelser2}
C.~Zeng and V.~Elser,
\newblock Phys. Rev. B \textbf{51}, (1995), 8318.

\bibitem{Sindzingre}
P.~Sindzingre, P.~Lecheminant and C.~Lhuillier,
\newblock Phys. Rev. B \textbf{50},  (1994), 3108.

\bibitem{Nakamura}
T.~Nakamura and S.~Miyashita,
\newblock Phys. Rev. B \textbf{52}, (1995), 9174.

\bibitem{Lecheminant}
P.~Lecheminant, B.~Bernu, C.~Lhuillier, L.~Pierre and P.~Sindzingre,
\newblock Phys. Rev. B \textbf{56}, (1997), 2521.

\bibitem{Fazekas}
P.~Fazekas and P.W.~Anderson,
\newblock Philos. Mag. \textbf{30}, (1974), 423.

\bibitem{Anderson}
P.W.~Anderson,
\newblock Science \textbf{235}, (1987), 1196.

\bibitem{Elser}
V.~Elser,
\newblock Phys. Rev. Lett. \textbf{62}, (1989), 2405.

\bibitem{Fisher}
M.~Fisher,
\newblock Phys. Rev. \textbf{124}, (1961), 1664.

\bibitem{Kasteleyn}
P.~Kasteleyn,
\newblock Physica \textbf{27}, (1961), 1209;
P.~Kasteleyn,
\newblock J. Math. Phys. \textbf{4}, (1963), 287.


\bibitem{Mila} F.~Mila, Phys. Rev. Lett. {\bf 81}, (1998), 2356.


\bibitem{remark} {In Ref \cite{Mila}, the model of Fig. \ref{fig:trimerized}
was called ``dimerized {\it kagom\'e}''. This terminology is misleading since
the local units consist of three sites. So we will use the more natural
terminology of ``trimerized {\it kagom\'e} model'' throughout.}

\bibitem{Sutherland}
B.~Sutherland,
\newblock Phys. Rev. B \textbf{37}, 3786 (1988).

\bibitem{remark1} {Note that the definition of $\vert \tilde{\psi} \rangle$
implies trivially that it is an eigenstate of each ${\cal S}_i$ with the
eigenvalue $\chi_{s} ({{\cal S}}_{i})$.}

\bibitem{remark2} {It is not immediately clear that what we call ${\text{deg}}(p_i)=
\sum_{q \in {\cal E}_i}
\chi_{s} ({{\cal S}}_{q})$ does not depend on the sector $s$ since it
involves sums over characters of this given sector. The argument is the
following: ${\text{deg}}(p_i)=\sum_{q \in {\cal E}_i}
\chi_{s} ({{\cal S}}_{q})= \sum_{q \in {\cal E}_i}
\chi_{s} ({{\cal S}}_{q} {{\cal S}}_{p_0})$ where $p_0$ is arbitrarily
chosen in ${\cal E}_i$. This last point is due to the fact that $\{ {{\cal
S}}_{q},\; q \in {{\cal E}_i} \}$ is a subgroup of $\{ {{\cal
S}}_{q} ,\;q=1,...,N_{\cal S} \}$. After a summation on $p_0$ over ${{\cal
E}_i}$ using the property $\chi_{s} ({{\cal S}}_{p} {{\cal S}}_{q}) =
\chi_{s} ({{\cal S}}_{p}) \chi_{s} ({{\cal S}}_{q})$ one gets
$({\text{deg}}(p_i))^2={\text{deg}}(p_i) {\text{Card}}({\cal E}_i)$
. Only two possibility can occur: either ${\text{deg}}(p_i)=0$ which is the case of
eliminated representatives, or ${\text{deg}}(p_i)={\text{Card}}({\cal
E}_i)$, which indeed does not depend on the sector $s$.}

\end{thebibliography}
\end{document}